# AGEM: Adaptive Greedy-Compass Energy-aware Multipath Routing Protocol for WMSNs

Samir Medjiah[*,**], Toufik Ahmed[**], Francine Krief[**]

*Ecole Nationale Supérieure d'Informatique(ESI),
Ex- Institut National d'Informatique (INI)
BP 68M Oued Smar,
16309 – Alger - Algérie
medjiah@ieee.com

**CNRS-LaBRI,
Université de Bordeaux-1.
351 Cours de la Libération,
33405 – Talence - France
{tad, krief}@labri.fr

*Abstract*—This paper presents an Adaptive Greedy-Compass Energy-aware Multipath (AGEM), a novel routing protocol for wireless multimedia sensors networks (WMSNs). AGEM uses sensors nodes position to make packet forwarding decisions. These decisions are made online, at each forwarding node in such a way that there is no need for global network topology knowledge and maintenance. AGEM routing protocol performs load-balancing to minimize energy consumption among nodes using twofold policy: (1) smart greedy forwarding based on adaptive compass and (2) walking back forwarding to avoid holes. Performances evaluations of AGEM compared to GPSR (Greedy Perimeter Stateless Routing) show that it can maximize the network lifetime, guarantee quality of service for video stream transmission, and scale better on densely deployed wireless sensors network.

## I. INTRODUCTION

With the growing-up of miniaturization technology and the availability of low-cost hardware, the sensors nodes embed nowadays various kinds of capturing elements such as microphones, imaging sensors, and video cameras. In this context, the vision of ubiquitous Wireless Multimedia Sensor Networks (WMSNs) [1][2][3] has become a reality. A sensor node gathers desired data information, processes it, and transmits it to each other using wireless communication until a base station. The base station (also referred to as the sink node) collects and analyzes the received data from various sensors and draws conclusions about the monitored area.

WMSNs are commonly used for surveillance applications, intrusion detection, environmental monitoring, etc. These types of applications require addressing additional challenges for energy-efficient multimedia processing, optimal routing and path selection, audio / video rate adaptation to meet the network changing topology, and application specific QoS guarantee.

Optimal routing in wireless sensor network is a challenging task. Large amounts of research works have been done to enable energy efficiency in WSN. A comprehensive survey of routing protocols in WSN has been presented in [4].

Routing protocols developed for WMSNs suggest using multipath selection scheme to maximize the throughput of streaming data. Examples of these protocols include: MPMPS (*Multi-Priority Multi-Path Selection*) [5] and TPGF (*Two-Phase Geographical Greedy Forwarding*) [6]. However, such protocols have to build a complete map of the network topology to select the optimum routing / transmission path between the source and the destination. They are not adapted in large-scale, high densely deployed network and frequent mobility situations.

Geographical routing can achieve scalability in WSNs. GPSR (*Greedy Perimeter Stateless Routing*) [7] was defined to increase network scalability under large number of nodes. The advantage is that the propagation of topology information is required only for a single hop. However, greedy forwarding relays on local-knowledge in which always best node to destination is selected. In such a case, selecting the same path using GPSR will lead to premature dying of nodes along this path.

In this paper, we examine the benefit of geographical routing along with multipath local-based route selection and we propose a new routing algorithm namely AGEM (an Adaptive Greedy-Compass Energy-Aware Multipath) routing protocol that leverages both energy constraint and QoS sensitive stream such as audio and video.

The design of AGEM was driven by the following points:

- *Shortest path transmission:* multimedia applications generally have a delay constraint which requires that the multimedia streaming in WSNs should always use the shortest routing path which has the minimum end-to-end transmission delay.
- *Multipath transmission:* Packets of multimedia stream are generally large in size and the transmission requirement can be several times higher than the maximum transmission capacity of sensor nodes.
- *Load balancing*: because of the density of a WSNs, a load balancing feature during the design of a routing protocol has to be considered to avoid frequent node failures and consequently to maximize the network lifetime.
- *Node selection*: in densely deployed network, different candidate neighbors may be used for packet forwarding. AGEM uses *adaptive compass* method to select candidate neighbor nodes which are in the line of sight towards the target the destination.

The rest of this paper is organized as follow. To make this paper self readable, we expose in section II the routing protocols that influenced the design of AGEM. In section III, we present the functionalities of AGEM protocol. In section IV, the performance evaluation of AGEM will be presented. Section V will conclude this paper.

## II. RELATED WORK

Geographic routing sheds light upon the process in which each node is aware of its geographic positing and uses packet's destination to perform routing decisions. In geographic routing, two greedy schemes are used to make packets progress towards the sink node. Greedy progression scheme based on distance to the sink node [7][8][9][10] and greedy progression based on angular offset from the direction towards the sink node [11][12][13]. In both schemes a path is dynamically constructed from the originating node to the destination using only local forwarding decisions.

For WMSNs, two important protocols have been defined that make use of node positing for packet forwarding decision: GPSR and MPMPS. MPMPS is itself based on TPGF. These protocols are briefly described in what follows.

### A. GPSR

The GPSR (*Greedy Perimeter Stateless Routing*) [7] was originally designed for MANETs but rapidly adapted for WSNs. The GPSR algorithm uses the location of nodes to forward a packet. It assumes that each node knows its geographic location and geographic information about its direct neighbors. This protocol uses two different packet forwarding strategies: *Greedy Forwarding* and *Perimeter Forwarding*. In *Greedy Forwarding* and when a node receives a packet destined to a certain node, it chooses the closest neighbor out-of itself to that destination and forwards it. Sometimes, such node cannot be found, (i.e. the node itself is the closest node to the destination out-of its neighbors), this situation is called a "void" or a "hole". Voids can occur due to random nodes deployment or the presence of obstacles that obstruct radio signals. To overcome this problem, *Perimeter Forwarding* is used to route packets around voids. Packets will move around the void until arriving to a node closest to the destination than the node which initiated the *Perimeter Forwarding*, after which the *Greedy Forwarding* takes over.

### B. TPGF

TPGF (*Two Phase geographical Greedy Forwarding*) [6] routing protocol is the first to introduce multipath concept in WMSNs field. This algorithm focuses in exploring and establishing the maximum number of disjoint paths to the destination in terms of the end-to-end transmission delay and the energy consumption of the nodes among a path.

The first phase of the algorithm explores the possible paths to the destination. A path to a destination is investigated by labeling neighbors nodes until the base station with a step back and mark feature to bypass voids and loops. The second phase is responsible for optimizing the discovered routing paths with the shortest transmission distance (least number of hops). The TPGF algorithm can be executed repeatedly to look for multiple node disjoint-paths.

### C. MPMPS

The MPMPS (*Multi-Priority Multi-Path Selection*) [5] protocol is an extension of TPGF. MPMPS highlights the fact that not every path found by TPGF can be used for transmitting video because a long routing path (i.e. long end-to-end transmission delay) may not be suitable for audio/video streaming. Furthermore, because in different applications, audio and video streams play different roles and the importance level may be different, it is better to split the video stream into two streams (video/image and audio). Therefore, we can give more priority to the important stream depending on the final application to guarantee the using of the suitable paths.

### D. Policies for Greedy forwarding

In literature, there are different policies that can be used in geographic routing and for the selection of the next hop node. To illustrate these policies, let take $u$ as the current forwarder node and $d$ the destination node (see Figure 1), then we can define these routing policies:

- **Compass routing:** The next relay node is $v$ such that the angle $\angle vud$ is the smallest among all neighbors of $u$ [11].
- **Random compass routing:** Let $v_1$ be the node above line ($ud$) such that $\angle v_1 ud$ is the smallest among all such neighbors of $u$. Similarly, define $v_2$ to be node below line ($ud$) that minimize the angle $\angle v_2 ud$. Then, node $u$ randomly chooses $v_1$ or $v_2$ to forward the packet [11].
- **Greedy routing:** The next relay node is $v$ such that the distance $\|vd\|$ is the smallest among all neighbors of $u$ [14][14].
- **Most forwarding routing (MFR):** The next relay node is $v$ such that $\|v'd\|$ is the smallest among all neighbors of $u$, where $v'$ is the projection of $v$ on segment $ud$ [15].
- **Nearest neighbor routing (NN):** Given a parameter angle $\alpha$, node $u$ finds the nearest node $v$ as forwarding node among all neighbors of $u$ in a given topology such that $\angle vud \leq \alpha$.
- **Farthest neighbor routing (FN):** Given a parameter angle $\alpha$, node $u$ finds the farthest node $v$ as forwarding node among all neighbors of $u$ in a given topology such that $\angle vud \leq \alpha$.
- **Greedy compass:** Node $u$ first finds the neighbors $v_1$ and $v_2$ such that $v_1$ forms the smallest counterclockwise angle $\angle duv_1$ and $v_2$ forms the smallest clockwise angle $\angle duv_2$ among all neighbors of $u$ with the segment $ud$. The packet is forwarded to the node of $\{v_1, v_2\}$ with minimum distance to $d$ [12][16].

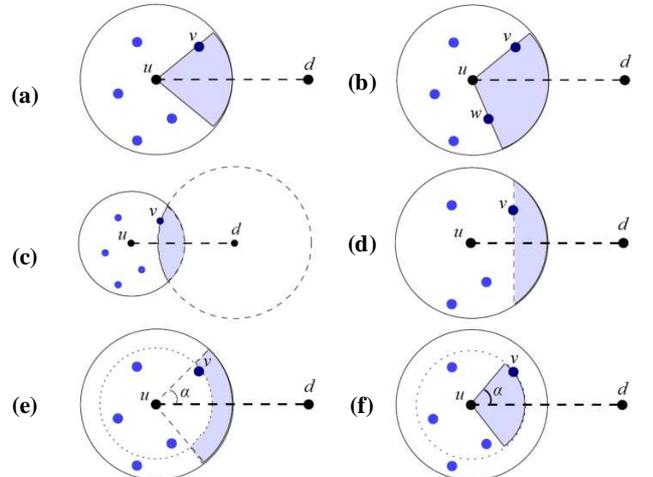

Figure 1: Different localized routing methods; (a) Compass, (b) Random compass, (c) Greedy, (d) Most forwarding, (e) further neighbor and (f) Nearest neighbor.

*E. Discussion*

Generally, a WSN is covered by densely deployed sensor nodes. Knowing the full map (network topology) of the deployed nodes in the network to perform routing as done by TPGF and MPMPS is not suitable for many reasons: (1) the exchange of the network map is energy consuming, (2) the exchanged map may not reflect the actual topology of the network, (3) nodes mobility and nodes failure are more frequent in WSN than in other ad hoc networks. These reasons are valid when paths are selected *a priori* by protocols such as TPGF and MPMPS. In such a case, the selected path is chosen in advance from the source to the destination based on route discovery mechanisms which run before the delivery phase. However, the actual map of the network may change. The GPSR protocol forwards the packet hop by hop based on local available information (Greedy routing policy). GPSR seems to be more promising to scale to large network but does not achieve load balancing in a statistical sense and by making use of multipath routing in WSNs.

In this paper, we propose a new geographical routing protocol namely AGEM (*Adaptive Greedy-Compass Energy-Aware Multipath*) that (1) selects neighbor nodes using adaptive compass mechanism which is considered as a new routing policy, (2) routes information on multipath basis using greedy routing functionalities and load balancing, and (3) avoids holes using walking back forwarding.

### III. AGEM ROUTING PROTOCOL

The AGEM routing protocol can be seen as an enhancement of the GPSR protocol to support the transmission of multimedia streams over WSNs by introducing adaptive greedy compass policy. The main idea is to add a load-balancing feature to GPSR in order to increase the lifetime of the network and to reduce the queue size of the most used nodes. In fact, routing data streams with GPSR will always choose the same path. This will rapidly cause the dying (dropping) of the most used nodes. In AGEM routing protocol, data streams will be routed by different paths.

At each hop, a forwarder node decides through which neighbor it will send the packet. Forwarding policy at each node is based on these four rules: (1) the remaining energy at each neighbor, (2) the number of hops made by the packet before it arrives at this node (3), the actual distance between the node and its neighbors, and (4) the history of the packets forwarded belonging to the same stream. Furthermore, only a subset of available neighbors is chosen according to adaptive compass policy.

The AGEM routing protocol has two modes, the *Smart Greedy Forwarding* and the *Walking Back Forwarding*. The first mode is used when there is always a neighbor closer to the sink node than the forwarder node, while the second one is used to get out of a blocking situation in which the forwarder node can no longer forward the packet towards the sink node. Figure 2 presents an overview diagram of AGEM routing mode switching. The following section will explain the two routing modes.

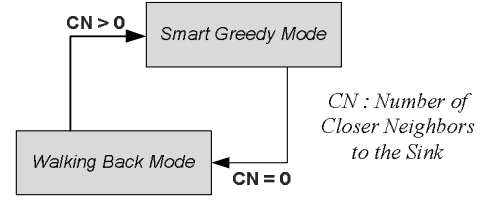

Figure 2: AGEM routing mode switching.

*A. Smart Greedy forwarding mode:*

The AGEM is a geographic routing protocol. Nodes are aware of their geographic coordinates. This information can be obtained using a positioning system such as GPS or by using a distributed localization techniques such as DV-Hop[17], Amorphous[18][19] …

In AGEM routing protocol, each sensor node stores some information about its *one*-hop neighbors. Information includes the estimated distance to its neighbors, the distance of the neighbor to the sink, the data-rate of the link, and the remaining energy. This information is updated by the mean of beacon messages, scheduled at fixed intervals. Relaying on this information, a forwarder node will give a score to each neighbor according to an objective function "*f(x)*".

Since AGEM protocol relays on beacon exchange for neighborhood maintenance, AGEM can be used for static sensor networks as well as for mobile sensor networks. The beaconing interval can be adjusted to meet the network dynamic.

*Best neighbor selection using adaptive compass policy*

AGEM relies on the basis of various possible forwarder neighbors towards the sink node. The AGEM routing algorithm includes an adaptive compass node selection (i.e. adaptive angle) which tries to select nodes with smallest angular offset from a virtual line toward the destination and satisfy a minimum number of nodes to ensure online multipath routing. Figure 3 illustrates the adaptive compass policy to meet the required number of forwarder nodes.

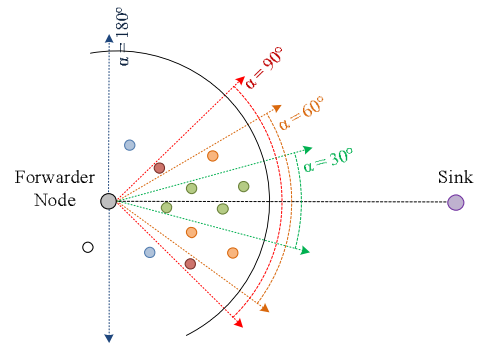

Figure 3: AGEM adaptive compass policy.

At the beginning, the forwarder node choose only neighbor nodes which are in the angle of view towards the destination with parameters α<30°. A minimum of two neighbor nodes (called neighbor set) must be found to perform load balancing, otherwise, the α angle is incremented by 10° until it reaches 180°. At this stage, if no nodes can be found and

α=180° then a walking back forwarding is needed since the forwarder is facing a hole.

Choosing a node to forward a packet among the neighbor set will depend on the score given to each node according to the objective function "*f(x)*". The *f(x)* considers the energy consumption with is defined in the following subsection.

*Packet energy consumption*

When a node ($A$) sends a packet ($pk$) of $n$ bits size to a node ($B$), the energy of node $A$ will decrease by $E_{TX}(n, \overline{AB})$ while the energy of the node $B$ will decrease by $E_{RX}(n)$. Consequently, the cost of this routing decision is $E_{TX}(n, \overline{AB}) + E_{RX}(n)$ considering the energy of the whole network.

We assume that the transmitted data packets in the network have the same size. We propose an objective function to evaluate a neighbor $N_i$ for packet forwarding. This objective function takes into account the packet energy consumption and also the initial energy of that neighbor. The proposed objective function can simply be:

$$f(N_i) = N_{i_{Energy}} - E_{TX}(N_{i_{Distance}}) - E_{RX}$$

Where: $E_{TX}(D)$ is the estimated energy to transmit a data packet through a distance D, and $E_{RX}$ is the estimated energy to receive the data packet.

These two functions rely on the energy consumption model proposed by *Heinzelman et al.* [20]. According to this model, we have:

$$E_{TX}(k, D) = k \cdot (E_{ELEC} + \varepsilon_{amp} \cdot D^2)$$
$$E_{RX}(k) = k \cdot E_{ELEC}$$

Where:

$k$ is the size of the data packet in bits,
$D$ is the transmission distance in meters,
$E_{ELEC}$ is the energy consumed by the transceiver electronics,
$\varepsilon_{amp}$ is the energy consumed by the transmitter amplifier.
$E_{ELEC}$ was taken to be 5 $\mu J/bit$ and $\varepsilon_{amp}$ 1 $\eta J/bit$.

For each known source node $s_i$ a forwarder node ($N$) maintains a couple ($H_i, j$). $H_i$ represents the mean hopcount that separates $s_i$ to $N$, and $j$ represent the neighbor whom score is closest to the average score of all closest nodes to the sink in the neighbor set (called best neighbor set). Since AGEM uses only an integer variable for each streaming source, any node can deal with multiple sources at one time and the memory requirements still reasonable for a sensor node.

Upon receiving a data packet from the source node $s_i$, the forwarder node will retransmit the packet to a neighbor that is closest to the sink node and in such a way that the number of hops the packet did, will meet the rank of that neighbor. The main idea is to forward a packet with the biggest number of hops through the best neighbor, consequently a packet with the smallest number of hops through the worst neighbor to allow best load balancing in the network. The following algorithm describes the forwarding policy.

Line 1 allows checking if we have already received a packet from a source node. If no, the packet will be always forwarded to the best node (line 2), and we have to save the hop count "H" and the average score index "j" in the best neighbor set. These empirical values will be used later to allow load balancing.

***Upon_Recieving_a_Packet*** ( *pk* )

*Inputs:*
**Best_Neighbor**: *a set of the closest neighbors to the sink node sorted in descending order by their score {$BN_1$, $BN_2$, … $BN_m$}.*
**m** = |**Best_Neighbor**|. **m** *represents the cardinal of the Best_Neighbor set*
**j** *:index of the node in the set **Best_Neighbor** whom score is closest to the average score of all closest nodes to the sink. For example, if **Best_Neighbor** is {8,5,2,1} the average score is **4** then **j=2** (starting from index=1)*

*Utilities:*
**Get_Hop_Values** ($S_i$) *returns the stored values of empirical hop count from already known source $S_i$ and the j index of the average score of all closest nodes to the sink. These values are ($H_i$, j)*
**Set_Hop_Values** ($S_i$, $H_i$, j) *sets the empirical hop count for source $S_i$ to be $H_i$ and j to be the index of the average score of Best_Neighbor set.*
**Forward** (**pk**, $BN_k$ ) *forwards the packet **pk** to the neighbor **k** which has $BN_k$ score*

| | |
|---|---|
| 1: | **if** (Get_Hop_Values (*pk*.SourceNode) **is** Null ) { |
| 2: |     Forward (*pk*, $BN_1$)        // *Default forward to best node* |
| 3: |     H ← *pk*.HopCount |
| 4: |     Set_ Hop_Values (*pk*.SourceNode, H, j) |
| | } |
| 5: | **else** {     //*Get_Hop_Values (pk.SourceNode) is not null* |
| 6: |     (H,j) ← Get_Hop_Values (*pk*.SourceNode) |
| 7: |     Δh ← H – *pk*.HopCount |
| 8: |     index ← j + Δh |
| 9: |     **case** (index ≤ 0) { |
| 10: |         H ← H–index +1 |
| 11: |         index←1 *// index of the best node in neighbor_Set* |
| 12: |     } |
| 13: |     **case** ( index > m ) { |
| 14: |         H ← H–index+m |
| 15: |         Index ←m *//index of the worst node in neighbor_Set* |
| 16: |     } |
| 17: |     Forward ( *pk*, $BN_{index}$ ) *// Smart forward* |
| 18: |     Set_ Hop_Values ( *pk*.SourceNode, H,j) |
| 19: | } |

**Figure 4: the Smart Greedy Forwarding algorithm.**

It is clear that the first packet received from an unknown source will be always forwarded to the best neighbor node.

Line 5 specifies that we have already an empirical estimation of the hop count H and the average index j from a particular source. These values are retrieved in line 6. We calculate in line 7, the deviation Δh of the hop count of the received packet compared to the stored value H. The index of the new forwarder neighbor that allows best load balancing will be adjusted by Δh (line 8). However, two different out of range situations may occur. Line 9 specifies that the received packet has experienced a lot of hops, and thus it needs to be forwarded later to the best node (i.e. node with index=1). In line 13, the received packet has experienced a less hop count than the empirical value H, and thus it has be forwarded to node with higher index (index=m). Line 10 and line 14 compute the new empirical value that will be used later as a new reference. Therefore, the smart forwarding occurs in line 17. Figure 5 illustrates this algorithm section.

### A. Walking Back forwarding mode

Because of node failure, node mobility, or scheduling policy, disconnections may occur in a WSN generating what we call "voids". In this situation, the neighbor set is empty and the angle α=180°. Thus, the forwarder node will inform all its neighbors that it cannot be considered to forward packets to the sink (see Figure 6). This node will also delegate the forwarding responsibility to the less far of its neighbors. This process is recursively repeated steps back until finding a node which can forward successfully the packet.

This technique is better than the perimeter routing mode used in GPSR routing protocol, since this kind of labeling mechanism is done only once when receiving a packet from an unknown stream, all the other packets belonging to the same stream will be routed avoiding the nodes that are facing a void toward the sink.

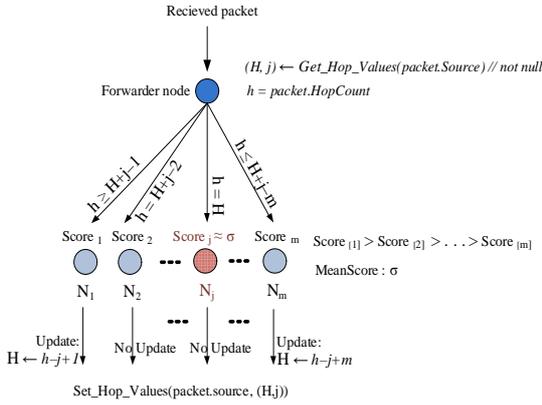

**Figure 5: Forwarding a packet of an already known stream.**

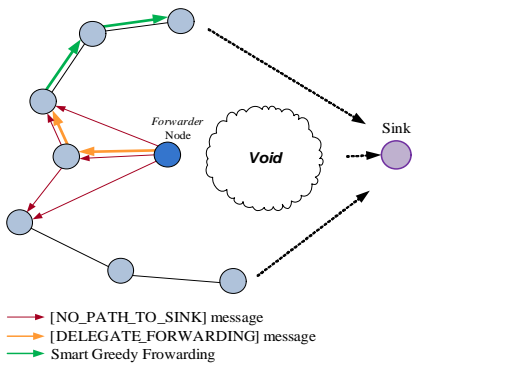

**Figure 6: A blocking situation where a forwarder node has no neighbor closer to the sink node than itself.**

### IV. SIMULATION AND EVALUATION

In this paper, we have considered a homogenous WMSN in which nodes are randomly deployed through the sensing field. The sensing field is a rectangular area of 500m x 200m. The sink node is situated at a fixed point in the righter edge of the sensing field at coordinates (490, 90) while a source node is placed in the other edge at coordinates (10, 90).

We consider this network for video surveillance. In response to an event, the source node will send images with a rate of 1 image per second during 30 seconds.

To demonstrate and evaluate the performance of AGEM, we used OMNeT++ 4.0 which is a discrete event network simulator [21]. To prove the effectiveness of AGEM, we have also implemented the GPSR algorithm and compared the simulation results. Table 1 summarizes the simulation environment.

We have considered that the link data is of type IEEE 802.15.4 and in which the data rate can be proportional to the transmission distance.

We have varied the network topology by varying the number of sensor nodes to obtain network of 30, 50, 80 and 100 nodes. We consider the minimum distance between two neighbors node greater than 1 meter.

| Parameter | Value |
|---|---|
| Network Size | 500m x 200m |
| Number of Sink Nodes | 1 |
| Number of Source Nodes | 1 |
| Number of Sensor Nodes | 30, 50, 80, 100 |
| Number of Images | 30 images |
| Image Size | 10Kb |
| Image Rate | 1 image/sec |
| Maximum Radio Range | 80 meters |
| Link Data Rate | $250\ Kbps/\sqrt{Link\_Length}$ |

**Table 1: Simulation parameters.**

For each topology and with a initial angle α=30°, we have measured various parameters: the distribution of the network remaining energy (mean in Figure 7 and variance in Figure 8), the distribution of the mean energy consumption by partitioning the network into regions of 40 meters width (see Figure 9 and Figure 10), the distribution of the packet end-to-end transmission delay in Figure 11, and finally the number of lost packets in Figure 12.

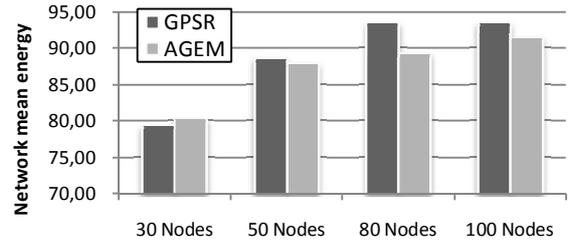

**Figure 7: The mean remaining energy in the different network topologies.**

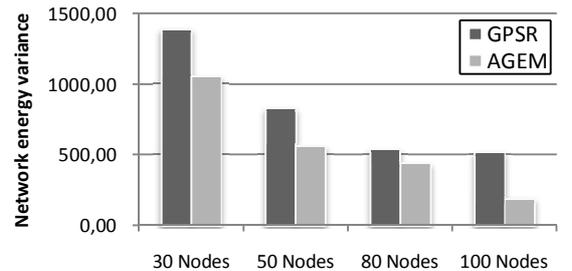

**Figure 8: The remaining energy distribution variance in the different network topologies.**

*Global energy distribution:*

Because of the inflexible selection of the next forwarder node, the GPSR keeps various nodes unused and utilizes a few nodes for sending packets. This explains that GPSR mean energy is being bigger than in the case of AGEM protocol as shown in Figure 7. However, the energy distribution in the network is well distributed with AGEM compared to GPSR, as illustrated in Figure 8, since most of the nodes can be active due to multipath routing.

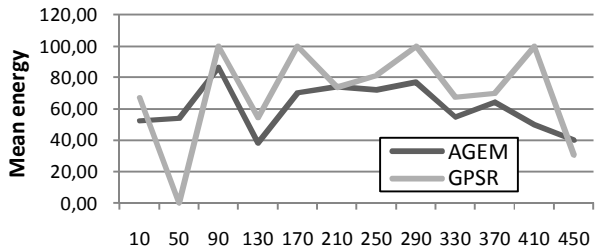

**Figure 9: The distribution of the remaining energy across the network for a 30 nodes network topology.**

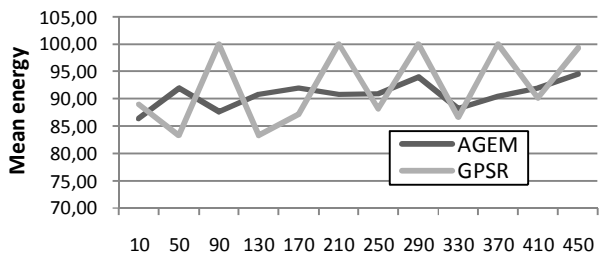

**Figure 10: The distribution of the remaining energy across the network for a 100 nodes network topology.**

*Local energy distribution:*

Figure 9 and Figure 10 illustrate the mean energy of the network partitioned in regions of 40 meters width for the topologies of 30 and 100 nodes. We can clearly see that the energy is uniformly consumed through the network when using AGEM routing protocol compared to GPSR routing protocol. The benefit of such a feature is preventing the network from being partitioned into sub networks completely disconnected if some nodes died before the others as we can clearly see in Figure 9 where no nodes remain alive in the region [50, 90] while using the GPSR algorithm.

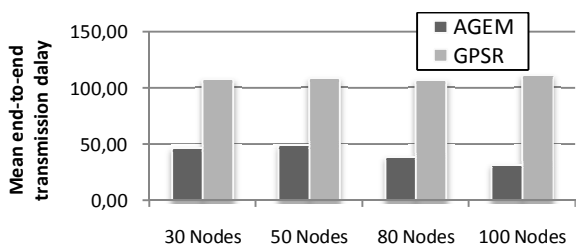

**Figure 11: The distribution of the end-to-end transmission delay.**

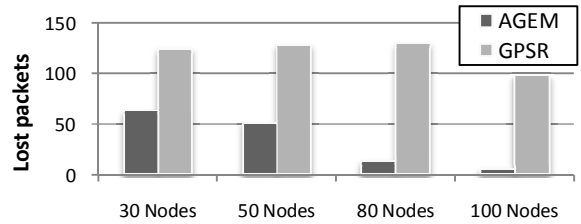

**Figure 12: The Number of lost packets.**

*Packet loss and Transmission delay:*

Because of the use of multiple paths to transmit data packets, the packet transmission delay has been extremely decreased as illustrated in Figure 11. The packet loss has also been decreased as shown in Figure 12. This enhancement can be explained by the following points:

− The use of the same path will increase the time spent inside the buffers (queue) among this path which leads to a traffic congestion.
− Packet loss may occur because sensors cannot keep packets for a long time in its buffers and this is due to the hard resource constraint.

These results demonstrate clearly the ability of AGEM to deliver multimedia traffic (Images traffic in our case) and enhancing the QoS compared to GPSR (lowering the end-to-end delay and packet loss ratio). AGEM is also more suitable to dense network in which different paths to destination may exist.

## V. CONCLUSION

In this paper, we have described a new algorithm namely AGEM that is suitable for transmitting multimedia streaming over WMSNs. Because nodes are often densely deployed, different paths from source nodes to the base station may exist. To meet the multimedia transmission constraints and to maximize the network lifetime, AGEM exploits the multipath capabilities of the WSN to make load balancing among nodes. Simulation results compared to GPSR show that AGEM is well suited for WMSNs since it ensures uniform energy consumption and meets the delay and packet loss constraint.